\documentclass[12pt]{article}
\usepackage{graphicx}
\usepackage{verbatim}

\input epsf
\def\beq{\begin{eqnarray}}
\def\eeq{\end{eqnarray}}

\def\ba{\begin{eqnarray}}
\def\ea{\end{eqnarray}}
\def\beq{\begin{eqnarray}}
\def\eeq{\end{eqnarray}}

\def\L*{{\cal L}_*}
\def\L{\mathcal{L}}
\def\({\left(}
\def\){\right)}

\def\lsim{\mathrel{\rlap{\lower3pt\hbox{\hskip0pt$\sim$}}
     \raise1pt\hbox{$<$}}}         
\def\gsim{\mathrel{\rlap{\lower4pt\hbox{\hskip1pt$\sim$}}
     \raise1pt\hbox{$>$}}}         

\def\lsim{\mathrel{\rlap{\lower3pt\hbox{\hskip0pt$\sim$}}
     \raise1pt\hbox{$<$}}}         
\def\gsim{\mathrel{\rlap{\lower4pt\hbox{\hskip1pt$\sim$}}
     \raise1pt\hbox{$>$}}}         

\usepackage{amsmath}
\usepackage{amsfonts}
\usepackage{verbatim}

\begin{document}

\begin{center}


{\Large\bf  Classicalize or not to Classicalize?}
 \vspace{0.2cm}

\end{center}

\begin{center}

{\bf Gia Dvali} 


\vspace{.2truecm}

\centerline{\em Arnold Sommerfeld Center for Theoretical Physics,
Fakult\"at f\"ur Physik} 
\centerline{\em Ludwig-Maximilians-Universit\"at M\"unchen,
Theresienstr.~37, 80333 M\"unchen, Germany}


{\em Max-Planck-Institut f\"ur Physik,
F\"ohringer Ring 6, 80805 M\"unchen, Germany}


{\em CERN,
Theory Division,
1211 Geneva 23, Switzerland}


{\em CCPP,
Department of Physics, New York University\\
4 Washington Place, New York, NY 10003, USA} \\

\end{center}



 

\centerline{\bf Abstract}

We show that  theories that  exhibit classicalization 
phenomenon cease to do so as soon as they  are endowed a Wilsonian weakly-coupled 
UV-completion that restores perturbative unitarity,   despite the fact  that such 
UV-completion does not change the leading structure of the effective low-energy theory. 
For example,  a Chiral Lagrangian of Nambu-Goldstone bosons (pions), with or without the 
Higgs (QCD)  UV-completion looks  the same in zero momentum limit, but  the latter  classicalizes in high energy scattering, whereas the former  does not.  Thus,  theory must  make a definite choice,   either  accept  a weakly-coupled UV-completion or  be  classicalized.  The UV-awareness that determines the choice is encoded in  sub-leading structure of effective  low-energy action. This peculiarity has to do with the fundamental fact that  in classicalizing theories high energies correspond to large distances, due to existence of the extended classical configurations sourced by energy.  UV-fate of the theory can be parameterized by introducing a concept of a new quantum length-scale,  de-classicalization radius.   Classicalization is abolished when this radius is a dominant length.   We then observe a possibility of a qualitatively new regime, in which a theory classicalizes 
only within a finite window of energies.  We suggest that one possible interpretation of physics  
above the classicality window is in terms of a  quantum theory of unstable extended objects.  In order to gain a physical intuition about the possible meaning of such regime,   
we establish analogy with  QCD-type theory (in which all quarks are heavy)   
which contains breakable QCD-type flux tubes.
 Extrapolating this analogy to light quark case, 
we observe that  QCD with light quarks can be regarded 
as a limit of  a would-be  classicalizing theory in which classicality window collapses to a single scale.


\section{Introduction}

  The focus of the present work is UV-completion of non-renormalizable theories in which 
 interaction strength  among  light fields is  controlled by positive powers  of a certain fundamental 
 length,  $L_*$.   Such interactions violate perturbative unitarity at energies above 
 $L_*^{-1}$. The standard (Wilsonian) approach to such theories is, that they can only be viewed as an effective  description of more fundamental weakly-coupled quantum theory valid at distances shorter than  $L_*$.  In this picture a weakly-coupled new physics 
 opens up at some intermediate distance $m^{-1}$, and restores perturbative unitarity. 
 
  It was suggested recently \cite{class1, class2}, that for a class of non-renormalizable theories, the picture of 
  UV-completion can be fundamentally different.   Namely, in such theories no weakly-coupled 
  new physics comes into the game.  Instead, at high-energies the system unitarizes itself by its own resources,   via {\it classicalization}.  That is, in such theories high-energy scattering amplitudes are dominated by 
  production of extended classical  configurations,  {\it classicalons}. 
  As a result, hard  $2\rightarrow 2$  scattering processes are exponentially suppressed, 
  and the scattering is  dominated by 
  \begin{equation}
  2 \, \rightarrow \, {\rm something ~classical}  \,  \rightarrow \, {\rm many } \, .
  \label{dominant}
  \end{equation}
   The essential feature of such theories   is the existence of a Bosonic degree of freedom, $\phi$, a so-called  {\it classicalizer}, which is  (self)sourced by energy.   Then,  at high enough center of mass energy, 
 $\sqrt{s} \, \gg \,  L*^{-1}$,  the scattering proceeds through the  formation of  an extended classical configuration of $\phi$ 
 of size $r_* \gg L_*$. Inevitability of such a configuration is due to the fact that $\phi$ is sourced by energy.  The effective size  of it sets the range of the interaction, and thus the cross section
 as,  
 \begin{equation} 
   \sigma  \sim  r_*^2 \, .
  \label{sigma}
  \end{equation}

 For better understanding  of classicalization phenomenon it is useful to  adopt an universal   
definition of the $r_*$-radius, given in \cite{class2}, which is equally applicable to  classicalizing as well as to  non-classicalizing theories.    According to this definition,  $r_*$ can be viewed as the length that at a given energy 
sets an effective range of the interaction.  That is, $r_*$ is a distance down to which 
wave-packets propagate essentially freely, without experiencing a significant scattering.    
In other words, the  equation (\ref{sigma}) can be used as definition of $r_*$ in an arbitrary theory. 
  Notice, that with this definition $r_*$  is automatically a classical scale, not containing 
  any powers of  $\hbar$ in it.   Notice also, that such a radius can be defined for an arbitrary quantum system, a familiar  example being a so-called classical radius of electron,
  $r_* \, = \, r_e \, \equiv \,  e^2/m_e$, that controls the interaction  range for Thomson scattering.  

$~~~$

  But, if $r_*$ can be universally-defined,  what determines  classicalization of a given theory? 
  
 $~~~$ 
  
   In the above language the answer is very simple.  The defining property is how the classical length $r_*$  relates to the relevant  quantum length-scales of the theory, and how this relation 
 evolves with energy.   
  In weakly-coupled theories $r_*$ is always much smaller than the quantum length-scales, and  
 with increasing $\sqrt{s}$ shrinks and becomes less and less relevant \cite{class2}.   For example, 
in Thomson scattering  $r_*$  is shorter than the electron Compton wave-length,  $r_* \ll m_e^{-1}$. 
 Because of this, by the time one is able to probe the distance $r_* = r_e$, the quantum effects become important.  Such systems are governed by quantum dynamics. 

  The condition for  classicalization is  the growth of $r_*$ with energy, in such a way, that 
 above some critical  energy it exceeds all the relevant quantum length scales in the system.  When this is the case, system 
 classicalizes.   This property is exhibited by derivatively -(self)coupled theories, in which 
 $\phi$ is sourced by energy. In such theories, depending on the concrete 
 interaction picture,  $r_*$ typically grows as a positive power of $\sqrt{s}$, 
 \begin{equation}
 r_*\sim L_*(L_*\sqrt{s})^{\alpha}   \, , 
 \label{stargr}
 \end{equation}
with $\alpha\, > \, 0$.  The precise value of $\alpha$ is determined by
a leading operator responsible for energy-sourcing, and is model-dependent.

 In this note we wish to further deepen the confrontation between the weakly-coupled and classicalizing theories, and show that the latter can be understood as a deformation of the former 
 in which the unitarizing weakly-coupled physics is removed, whereas the leading structure 
 of high-dimensional derivative-couplings of  low energy effective theory is kept in tact. 
 
  This transition from quantum-weakly-coupled to classicalizing behavior  
demonstrates explicitly that classicalization is a "self-defense" of a theory when it is deprived    
of a  weakly-coupled UV-completion.   Correspondingly, by  integrating-in a weakly-coupled physics 
that restores perturbative unitarity we {\it de-classicalized} the theory.

  We shall arrive to this conclusion by addressing the following (seeming) puzzle. 
 Consider a derivatively-self-coupled low energy theory,  which above a scale $m \, <  \, M_*\equiv L_*^{-1}$ admits a weakly-coupled completion.  The role of such theory,  for example, can be played by 
a chiral Lagrangian of a massless Goldstone field, $\phi$,  which at high energies gets completed as a theory of a weakly-coupled 
scalar  field  (Higgs)  that spontaneously breaks a continuous global symmetry. 
  
  Integrating out a radial (Higgs) degree of freedom, we get an effective low energy 
theory which contains derivative interations of the Nambu-Goldstone  bosons.  
 As was argued in \cite{class1, class2}, such a theory classicalizes at 
 a distance $r_* \, \gg \, L_*$.   Notice, that classicalization is  a
 high-energy, but {\it long-distance} effect.   But, at long distances, seemingly, theory should know nothing about  its perturbative  UV-completion.   Thus, theory should classicalize regardless of the existence of such a completion. But, this is impossible, since, as viewed from  high-energy perspective  theory is a weakly coupled theory of a scalar, which cannot exhibit any classicalization. 
  Weakly-coupled theories cannot  classicalize.  Since the two UV-behaviors are very different, one is left with a puzzle. 
  
  $~~~$
  
  How can an effective low-energy  theory decide which UV-path to choose:  {\it  Classicalize 
  or not to Classicalize?} 

$~~~$
 
   In this note we shall  resolve this seeming puzzle and show, that classicalization and weakly-coupled completions are inter-exclusive.   That is,  as long as  asymptotic degrees of freedom are fixed,  in any  given energy range, theory either classicalizes or has a weakly-coupled UV-completion, but not both properties simultaneously. 
   
  (Notice, that this  statement does not exclude possible duality-type relation between the classicalizing  and weakly-coupled descriptions, which changes the definitions of asymptotic  states.  Such possibility is  not our focus at the moment, but will become later when we go to QCD analogy, see below.)

    By integrating-in any weakly-coupled physics that restores 
   unitarity we  kill classicalization,  and theory de-classicalizes. 
  From the first glance,  this may come as a surprise, since as argued before, classicalization is a long-distance effect, since it takes place at distances  much lager than the cutoff length of the 
  low energy theory, $r_* \, \gg \, L_*$.   However, the key point is, that  in clasicalizing theories 
  the long distances are not necessarily equivalent to low-energies, due to existence 
  of extended objects, classicalons, that dominate high-energy scattering.  As a result, the 
  scale $r_*$  corresponds to the energies at which the theory 
  is already well-aware of the existence of weakly-coupled unitarizing physics, and 
  classicalons are never formed.  The theory de-classicalzes. 
  
   As we shall see, the decision about which path to follow is made by the operators that 
  are sub-leading at low-energies, but which carry information about the existence (or non-existence) of  unitarity-restoring (weakly-coupled)  propagating degrees of freedom in UV.    
  System classicalizes when such  information is absent.  
     
       Analyzing a possible influence of such  "UV-informative" operators on the classicalization dynamics, we  shall discover a possibility of qualitatively 
   new regime, in which the system classicalizes  within the finite interval of energies. 
  We shall refer to this energy interval  as  the {\it window of classicality}.   
    In such theories, at increasing center of mass energy, the system swhitches from
  quantum to classical and back to quantum regimes.  The space of such theories can be 
  parameterized by introducing a concept of  a new quantum length-scale, $\bar{r}$, which we shall refer to as the de-classicalization  radius.  This radius encodes information about the importance 
  of the quantum effects on the system and diminishes in the limit $\hbar \, \rightarrow \, 0$. 
     In our parameterization,  classicality window is determined as the range of energies 
     for which
     \begin{equation}
     r_* \, \gg \, \bar{r}, \, L_* \, .
     \label{cw} 
     \end{equation}
     Outside the classicality window one of the quantum scales (either $L_*$ or $\bar{r}$) starts to dominate over $r_*$, and system becomes quantum. 
   In this parameterization, clasicalizing theories correspond to the case when condition 
   (\ref{cw}) is satisfied in arbitrarily deep UV, and classicality window is infinite.    
  
   Given the required structure of the low energy theory, the finiteness of 
  classicality window can be detected unambiguously.   This rises an interesting question 
  about the  physical meaning of theories with finite classicality window.   Our analysis shows,  that beyond  this window in deep-UV theory  becomes again quantum, but  what are the corresponding degrees of freedom? 
  
    We speculate in this direction and suggest, that the UV-theory above classicality window  is a quantum theory of unstable extended objects.   In order to substantiate this guess, we use physical intuition coming from QCD-type theories in which all quarks are heavier than the QCD scale, 
   $ \Lambda_{QCD}$. Such theories contain flux-tubes (QCD-strings) that are practically-stable below certain critical (exponentially-long) size ($\equiv \, \bar{L}$), 
 whereas the longer tubes are unstable.  
  We show, that for such a system the complete analogs of  $L_*$, $r_*$ and $\bar{r}$ lengths can be defined, in terms of 
    QCD-confinement length ($L_{QCD} \, \equiv \, \Lambda_{QCD}^{-1}$),  classical length of the QCD flux-tube ($L$),
  and the ratio of the latter length-squared to the critical instability length ($L^2/\bar{L}$),  
     respectively. 
   This  analogy however has obvious limitations,  since such a QCD is a theory with a mass gap
   and contains no light analog of the classicalizer  field $\phi$.    In order to create such an analog, in form of pions, we need to push the quark masses below the QCD-scale. 
    But, such a deformation of the theory eliminates the classicality window, since 
   QCD flux-tubes longer than $L_{QCD}$  become unstable.  
    This gives a complementary  understanding of why, due to the presence of the heavier resonances,  the pion Lagrangian does not classicalize at high energies.  
 
   Thus, in classicalization language, the  low-energy  QCD can be viewed  as a (would-be classicalizing) theory  in which classicality window collapses towards the  QCD-scale.

   \section{Classicalization and  De-Classicalization}
   
     Before going to more detailed analysis, let us explain the main reason behind the de-classicalization phenomenon. 
    
    For this, let us first briefly summarize the essence of  classicalization effect.  The latter phenomenon is exhibited by  a class of theories with derivative self-interactions. 
  As a simple prototype example,   consider a theory of a real scalar field,  
   \begin{equation}
 \mathcal{L}= {1\over 2}  \(\partial_{\mu}\phi\)^2 \, + \, {L_*^4\over 4}  \((\partial_{\mu}\phi)^2\)^2 \, .
 \label{nambu}
  \end{equation}
  This theory is symmetric  under the shift by an arbitrary constant $c$,   
  \begin{equation}
  \phi \, \rightarrow \, \phi \, + \,  c \, . 
\label{goldstone}
\end{equation}  
 As it was shown in \cite{class1,class2}, this theory classicalizes at $\sqrt{s} \, \gg\, 1/L_*$.
 Namely, at high center of mass energies,  the scattering takes place at a macroscopic 
 length-scale determined by the  classical radius,  
 \begin{equation}
r_* \equiv  L_* (\sqrt{s} L_*)^{1 \over 3} \, .
 \label{1radstar} 
 \end{equation}
In order to have an unified description of classicalizing and non-classicalizing theories, it is useful
to adopt the definition of $r_*$ radius given in\cite{class2}. According to this definition, in a generic  theory 
the physical meaning  of $r_*$-radius is of a distance at which the scattering of a free wave becomes significant.
 Assuming  that for $r \, = \, \infty$ and $t  \, = \, - \infty$, 
$\phi$ starts out as an in-wave $\phi_0$  satisfying the free-equation,  
  \begin{equation}
  \Box\phi_0 \,   =  \, 0  \, ,  
     \label{equfree}
  \end{equation}
  and representing  $\phi$ as a superposition of free and scattered waves,
\begin{equation}
\phi \, = \, \phi_0 \, + \, \phi_1 \, , 
\label{zeroone}
\end{equation}
$r_*$-radius is defined as a characteristic distance  at which  the correction $\phi_1$ to a free-wave becomes 
significant.   That is,  $r_*$-radius is determined from the condition,  
\begin{equation}
\phi_0  (r_*) \, \sim \phi_1  (r_*) \, .  
\label{condi}
\end{equation} 

 Let us  now imagine, that we UV-complete the above theory by a weakly-coupled physics 
  at some scale  $m$. Obviously, in order to restore perturbative unitarity, the scale 
  $m$ at which the new physics sets-in must be sufficiently below the unitarity-violation scale $L_*^{-1}$.   In order to make the argument clean,  let us take, 
  \begin{equation}
  m \ll L_*^{-1} \,. 
  \label{hierarchy}
  \end{equation}
  The effect of such weakly-coupled UV-completion 
on the effective low energy action (\ref{nambu})  is expressed in  endowing 
an each additional square of the derivative  by 
a propagator with a massive pole at $m^2$,   
\begin{equation}
(\partial_{\mu}\phi)^2 \, \rightarrow \,  {m^2 \over m^2 \, +\  \Box} 
(\partial_{\mu}\phi)^2
\, , 
\label{replace}
\end{equation}
so that the low-energy effective action becomes
    \begin{equation}
 \mathcal{L} \, = \, {1\over 2}  \(\partial_{\mu}\phi\)^2 \, + \, {L_*^4\over 4}  \((\partial_{\mu}\phi)^2
  {m^2 \over m^2 \, +\  \Box}  (\partial_{\nu}\phi)^2 \)  \, + \,  ... \, .
 \label{nambuweak}
  \end{equation} 
  The qualitative reason of  why weakly-coupled Wilsonian UV-completion de-classicalizes  
the theory is now clear.   Although,  at low energies   ($\Box \, \ll \, m^2$) the action 
(\ref{nambuweak}) effectively reduces to (\ref{nambu}), the difference is substantial 
at high energies  ($\Box \, > \, m^2$).  Above this energy, the mass term in the 
propagator becomes irrelevant and the interaction becomes suppressed by momentum-squared.   

 As a result of this suppression,  in  high center of mass scattering,  
 the self-sourcing by energy switches off at distances much larger than the would-be 
 classicalization radius, and waves scatter  without ever reaching the classicalization point.  
  Theory de-classicalizes. 

 Indeed in the absence of weakly-coupled UV-completion,  as in theory (\ref{nambu}),  
 at center of mass energy $\sqrt{s} \, \gg \, L_*^{-1}$ classicalization  would happen at distance 
 $r_* \gg L_*$,  for which 
 \begin{equation}
   \Box \phi_0^2 \, \sim \, L_*^{-4}  \, .
 \label{claspoint}
 \end{equation}
 Notice,  that for any free wave satisfying (\ref{equfree}), the above condition is equivalent to 
 \begin{equation}
  ( \partial_{\mu} \phi_0)^2 \, \sim \, L_*^{-4}  \, .
 \label{claspoint2}
 \end{equation}
  For classicaliozation  it is absolutely  essential that $r_*$-radius for which 
  the equivalent conditions (\ref{claspoint}), (\ref{claspoint2}) and (\ref{condi}) are satisfied, 
  is much larger than the fundamental (quantum) length $L_*$.   And in theory (\ref{nambu}) this is indeed the case, according to (\ref{1radstar}).

  It is now obvious that in a weakly-coupled completion this situation  can never be reached, since 
 the distance at which the self-sourcing switches off is given by the radius $\bar{r}$ at which
  \begin{equation}
   \Box \phi_0^2 \, \sim \, m^2 \phi_0^2  \, . 
 \label{sourceoff}
 \end{equation}
This  gives 
  \begin{equation}
   \bar{r} \, \sim \, \sqrt{s}/m^2 \, . 
 \label{rd}
 \end{equation}
Obviously, for $m \ll L_*^{-1}$, we have $\bar{r} \,  \gg \, r_*$ and thus system never reaches the classicalization radius. 
 
  We shall now discuss the above effect in more details.


  \section{Classicalons}
  
  As an example of classicalizing theory,  consider again the prototype  model of 
  derivatively self-interacting scalar, given by the action (\ref{nambu}).  
 In order to see why the theory classicalizes  at energies $\sqrt{s} \, \gg\, 1/L_*$,  let us repeat the steps of \cite{class2}. 
   For this we shall study the scattering of waves by analyzing  the equation of motion
 following from (\ref{nambu}),   
  \begin{equation}
  \partial^{\mu} (\partial_{\mu}\phi \(1+ L^4_*(\partial_{\nu}\phi)^2)\) = 0 \, . 
  \label{goldequation}
  \end{equation}
We shall assume that for $r \, = \, \infty$ and $t  \, = \, - \infty$, 
$\phi$ is well-approximated by a spherical wave of characteristic frequency $\omega$ and 
the amplitude $A \sim 1$ given by 
\begin{equation}
\phi_0 \, = \, {\psi(\omega(r+t))  \over r},
\label{wavepacketfree}
\end{equation}
which solves the free-field equation of motion (\ref{equfree}).  
Notice, that for small occupation number $A\sim 1$,   the energy of the wave is $\sqrt{s} \sim \omega$. 

We shall now solve the equation (\ref{goldequation})  iteratively, by representing the $\phi$-field
according to (\ref{zeroone}) as  a superposition of a free-wave $\phi_0$ and a scattered wave $\phi_1$. We shall treat $\phi_1$ as a small correction, 
and try to understand at what distances it  becomes 
significant.   The $r_*$-radius is then determined  by a distance for which,  $\phi_1$ becomes comparable to $\phi_0$, according to (\ref{condi}).  For our purposes we do not need to go beyond this point.  
  In the leading approximation, the equation for 
the correction to the free wave is, 
 \begin{equation}
  \Box\phi_1  =  - L^4_*
     \partial^{\mu} (\partial_{\mu}\phi_0 (\partial_{\nu}\phi_0)^2) \, .
     \label{phione1}
  \end{equation}
Taking into the account properties of $\psi(\omega(t+r))$-wave,  for 
$\omega \, \gg \, r$, the leading contribution to the right hand side is,  
 \begin{equation}
  \Box\phi_1  =  -  {L^4_*  \over r^5} \, ( 2 \psi^2\psi '' \, + \, 8 \psi\psi '^2) \, ,
     \label{phionea}
  \end{equation}
where prime denotes the derivative with respect to the argument. 
For $\omega \, \gg \, r^{-1}$ the solution of this equation can be approximated by, 
  \begin{equation}
  \phi_1 \, \simeq \, - f(\omega(r+t)) \,  \frac{L_*^4 }{6 r^{4}} \, ,  
  \label{aprphi1}
  \end{equation}
  where,  
 \begin{equation}
 f(\omega(r+t)) \, \equiv  \, \int_0^{r+t} ( 2 \psi^2\psi '' \, + \, 8 \psi\psi '^2) dy \, .
 \label{fdefinition}
 \end{equation}  
 Notice, that  since  $\psi(wy)$ is a bounded function of amplitude $\sim 1$ and characteristic frequency  $\omega$,   we have,   
 \begin{equation}
f \, \sim \,  \,  \omega \psi \, \sim \, \omega \, . 
  \label{omegaplus}
  \end{equation}
  Thus,  we obtain the following relation between the initial free-wave and the leading perturbation due to scattering, 
  \begin{equation}
  \phi_1 \, \sim \, \left ({r_* \over  r} \right )^3 \phi_0 \, ,
  \label{starrelation}
  \end{equation}
where, 
\begin{equation}
r_* \equiv  L_* (\omega L_*)^{1 \over 3} \, .
 \label{radstar} 
 \end{equation}
 This equation tells us, that  the scattering of Goldstones starts 
already  at a distance $r_*$, which grows at large  $\omega$, as suggested originally in \cite{class1}.
That is, the classical interaction range exceeds all the quantum length-scales in the problem. 
   This is the signal for formation of a classical configuration. 
  
  \section{De-Classicalization by Weakly-Coupled UV-completion}
  
 We shall now illustrate how weakly-coupled UV-completion  de-classicalizes the above theory. 
 Example of such UV-completion is given by an embedding of $\phi$ in a theory of a weakly-coupled complex scalar
 \begin{equation}
 \Phi \, \equiv \, {1 \over \sqrt{2}} (v  \, + \, \rho) {\rm e}^{i \phi/ v} \, 
 \label{complex}
 \end{equation}
 that spontaneously breaks  a  global $U(1)$-symmetry,  $\phi \rightarrow \phi + c$, 
 through its vacuum  expectation value  $\langle \Phi^*\Phi \rangle \, = \,  {1\over 2} v^2$. 
 Generalization to higher symmetries is straightforward.  
  Here $\phi$ and $\rho$ are  angular (Nambu-Goldstone)  and radial (Higgs) degrees of freedom respectively. 
 
 The microscopic theory has the following Langangian, 
    \begin{equation}
 \mathcal{L} \, = \,   |\partial_{\mu}\Phi|^2 \, - \,{\lambda^2 \over 8} \( 2 |\Phi|^2\, - \, v^2 \)^2 \,. 
 \label{weakcomplex}
  \end{equation}
 Or written in terms of Goldstone  and Higgs degrees of freedom,  
   \begin{equation}
 \mathcal{L}= {1\over 2} \left(1 \, + \,  2 \rho v^{-1}  \, + (\rho v^{-1})^2 \right) \(\partial_{\mu}\phi\)^2 \,  + \,  {1\over 2}  \(\partial_{\mu}\rho\)^2 
 \, - \,\left ( {m^2 \over 2} \rho^2\, + \, 
{ \lambda \over 2}  m \rho^3  \,  +    {\lambda^2 \over 8}  \rho^4 \right ) \, .
 \label{weak}
  \end{equation}
 This theory describes a massless Goldstone field $\phi$, coupled to a radial mode $\rho$ of mass
 $m \equiv \lambda v$.  
 
  Obviously,  this theory is a weakly-coupled theory and as such does not classicalize. 
  The $r_*$-radius at high energies shrinks as $r_* \sim \lambda /\omega$ \cite{class2}. 
  We now wish to understand this de-clasicalization  from the point of view of the low-energy theory 
  and show that it can be understood as an effect of integrating-in in theory (\ref{nambu}) 
  a weakly-coupled radial mode $\rho$. 
  
   For this we shall ingrate out the weakly-coupled heavy degree of freedom, $\rho$,  and write down an effective interaction for $\phi$.    
   We shall work with equations of motion which take the following form, 
  \begin{equation}
  (\Box \, + \, m^2) \rho\,   =  \,  \left ( v^{-1}  \, + \, \rho v^{-1} \right) \(\partial_{\mu}\phi\)^2 
  \, - {3\over 2} \lambda m \rho^2  -  {\lambda^2  \over 2} \, \rho^3 \, , 
     \label{equphi4mass}
  \end{equation}
  \begin{equation}
  \partial^{\mu} \left (\partial_{\mu}\phi \(1+  2 \rho v^{-1}  \, + (\rho v^{-1})^2 )\) \right )\,  = \, 0 \, . 
  \label{goldequationhiggs}
  \end{equation}
   Integrating out $\rho$  to the leading order  in  ${v^{-2} \over  \Box \, + \, m^2} \(\partial_{\mu}\phi\)^2 $, gives  the following effective equation of motion for $\phi$. 
  \begin{equation}
  \partial^{\mu} \left (\partial_{\mu}\phi \(1\, + \,  2 {L_*^4 m^2 \over  \Box \, + \, m^2} \(\partial_{\mu}\phi\)^2    \) \right ) \,  = \, 0 \, . 
  \label{goldequation1}
  \end{equation}
  where $L_* ^2 \, \equiv \, (mv)^{-1}$. 
 As we see, this equation of motion differs  from  (\ref{goldequation}) by the replacement 
\begin{equation}
L_*^4 \, \rightarrow  \,  {L_*^4 m^2  \over m^2 \, +\  \Box} \, , 
\label{replace2}
\end{equation}
 which is equivalent to  (\ref{replace}).
     In view of this, it is 
 not surprising that the corrected effective theory no longer clissicalizes.  
  We shall demonstrate this explicitly.  In order  to see this,  we shall consider the same scattering problem as above, preparing $\phi_0$  in form of a collapsing wave of small occupation number and 
  a high center of mass energy $\sqrt{s} = \omega$, and solve for the scattered wave    $\phi_1$ 
  in the leading order. The equation for $\phi_1$ now gets corrected by a massive propagator and
  after taking into the account  (\ref{equfree})  can be written as,  
   \begin{equation}
  \Box\phi_1  =  - 2L^2_* \, 
    (\partial_{\mu}\phi_0)  {1 \over m^2 \, +\  \Box}  \partial^{\mu}  (\partial_{\nu}\phi_0)^2 \, .
     \label{phicorrected}
  \end{equation}
 Since in the absence of weakly-coupled UV-completion, the classicalization is 
  taking place at distances for which $\Box \phi_0^2 \sim L_*^4$, and since 
  $m \ll L_*^{-1}$, for the distances of interest, we can ignore $m^2$ in the propagator, and the equation simplifies to, 
    \begin{equation}
  \Box\phi_1  =  - 2L^2_* \, 
    (\partial_{\mu}\phi_0)  {1  \over  \Box}  \partial^{\mu}  (\partial_{\nu}\phi_0)^2 \, .
     \label{phicorrected1}
  \end{equation}
   The latter equation can be easily integrated, noting,   that any free wave, satisfying $\Box \phi_0\, = \, 0$, 
  obeys the following relations,. 
   \begin{equation}
  \Box\phi_0^2  =  2 \, 
    (\partial_{\mu}\phi_0)^2 \, ~~~~  \Box\phi_0^3  =  3 \, 
   ( \partial_{\mu}\phi_0) (\partial_{\mu}\phi_0^2)  \, ,
     \label{phicorrected1}
  \end{equation}
 which enable to rewrite (\ref{phicorrected1}) in the following form, 
     \begin{equation}
  \Box\phi_1  =  -  {L^2_* \over 3}  \, \Box \phi_0^3 \, .
     \label{phicorrected2}
  \end{equation}
Thus, we arrive to the relation,  
     \begin{equation}
  \phi_1  =  -  {L^2_* \over 3}  \phi_0^3 \, ,
     \label{phicorrected3}
  \end{equation}
which implies that classicalization never happens, since $r_*$ shrinks below $L_*$. Of course, 
for $r \sim L_*$ one has to consider subleading correction, or even better simply work in   
high-energy formulation in terms of weakly-coupled complex scalar $\Phi$. This formulation 
of course suggests that $r_*$ shrinks as  $r_* \sim \lambda/\omega$, so that 
it becomes shorter than any relevant quantum scale in the theory. 
  In other words,  system de-classicalizes.  
  
  \section{Generalization to an Arbitrary Theory}
 
   Our analysis  showing that a weakly-coupled UV-completion de-classicalizes the theory
   can be easily generalized to  an arbitrary theory in which perturbative unitarity is restored 
   by weakly-coupled physics.   Indeed,  in classicalizing theories  an  effective would-be perturbative coupling in a given vertex blows up  with increasing energy.   This blow-up is the key 
   both for violation of perturbative unitarity as well as  for classicalization.   Any weakly-coupled 
   UV-completion that shuts-off this growth and restores perturbative unitarity must  automatically  
  kill classicalization.  
   
     For example, consider a generalization of our previous example to  other higher order invariants, by  adding derivative couplings of the form, 
 \begin{equation}
 \mathcal{L}= {1\over 2}  \(\partial_{\mu}\phi\)^2 \, + \, \sum_{n>1} {a_nL_*^{4(n-1)}\over 2n}  \((\partial_{\mu}\phi)^2\)^n \, , 
 \label{nambuN}
  \end{equation}
 where $a_n$ are some constant coefficient.  For example,   the  Dirac-Born-Infeld Lagrangian, 
 \begin{equation}
L_{DBI} \, = \,  L_*^{-4}\sqrt{1 \, + \,  L_*^4(\partial_{\mu}\phi)^2 } \, ,
 \label{DBI}
 \end{equation}
 considered as one of the classicalizing examples in  \cite{class1},  corresponds to a particular choice of  coefficients $a_n$.   
 
 The equation of motion has the following form,  
 \begin{equation}
 \partial^{\mu} \left (\partial_{\mu}\phi \(1 \, + \, \sum_{n>1} a_nL_*^{4(n-1)}  \((\partial_{\mu}\phi)^2\)^{n-1} \) \right ) \, =  \, 0 \, . 
 \label{nambuN}
  \end{equation}

  For an incoming free wave-packet 
 $\phi_0$, the equation describing a scattered wave $\phi_1$  for such a theory is 
  \begin{equation}
 \Box \phi_1 \, = - \partial^{\mu} \left (\partial_{\mu}\phi_0\,  \sum_{n>1} a_nL_*^{4(n-1)}  \((\partial_{\mu}\phi_0)^2\)^{n-1} \right ) , 
 \label{nambuN1}
  \end{equation}
which implies that 
  \begin{equation}
  \phi_1 \, \sim \,  {1 \over r}  \sum_{n>1} a_n\left ({\omega \over r^3}  L_*^{4}\right )^{n-1} \,. 
 \label{phi1N}
  \end{equation}
This gives the  classicalization radius  $r_* \, \sim \, L_* (L_*\omega)^{1/3}$.  
  
   Integration-in of a weakly-coupled physics at scale $m$  is  expressed in modification 
   in which  each $a_n$-th term  in (\ref{nambuN}) is replaced by  combinations of the form
\begin{equation}  
 {m^2 \over m^2 \, +\  \Box} \left ( (\partial_{\nu}\phi)^2 {m^2 \over m^2 \, +\  \Box} \left(  (\partial_{\nu}\phi)^2  {m^2 \over m^2 \, +\  \Box} \, ... (\partial_{\nu}\phi)^2  {m^2 \over m^2 \, +\  \Box}  (\partial_{\nu}\phi)^2 \right)... \right) \, .
   \end{equation}
The perturbative restoration of unitarity requires that each  $(\partial_{\nu}\phi)^2$ is accompanied 
by at least one propagator ${m^2 \over m^2 \, +\  \Box}$,  so that 
the combination $({m^2 \over m^2 \, +\  \Box}  (\partial_{\nu}\phi)^2)$ appears $n-1$-times. 

 Then at short distances ($\Box \, \gg \, m^2$), the equation for the scattered wave becomes      
 \begin{equation}
 \Box \phi_1 \, = - \partial^{\mu} \left (\partial_{\mu}\phi_0\,  \sum_{n>1} a_nm^{2(n-1)}L_*^{4(n-1)}  \left (
 {1 \over  \Box} \left ( (\partial_{\nu}\phi_0)^2 {1 \over  \Box} \left(  (\partial_{\nu}\phi_0)^2  {1 \over  \Box}  \left ( 
 ... {1 \over  \Box}  (\partial_{\nu}\phi_0)^2\right )\,...\, \right)\right) \right) \right)
 \label{nambuN1}
  \end{equation}

  This seemingly complicated equation is trivially integrated, by using the identity 
  $\Box \phi_0^n = n(n-1) \phi_0^{n-2}\,  (\partial_{\mu} \phi_0)^2 $. The result is 
   \begin{equation}
   \phi_1 \, = \,     \sum_{n>1} a_nm^{2(n-1)}L_*^{4(n-1)} \phi_0^{2n-1} \, ,
  \label{final}
  \end{equation}
  where all combinatoric factors have been absorbed in redefinition of parameters $a_n$, whose values are unimportant for our consideration anyway. What is crucial from the above expression 
  is,  that $r_*$ no longer grows with $\omega$, but rather shrinks at least to $L_*$;
  The system de-classicalizes.

   \section{Window of Classicality}
   
    We have seen,  that  for de-classicalization it is necessary that {\it every} additional  
 square of the derivative  in  the self-coupling gets accompanied by a regulating  factor 
(\ref{replace}), which suppresses the derivative-dependence at high momenta 
and switches off the energy self-sourcing.   That is,  every interaction term 
containing $2(n +1)$ powers of  derivatives,  must contain at least $n$ massive pole insertions. 
Schematically, we can write, 
 \begin{equation}
\partial^{2(n+1)} \, \rightarrow  \, \partial^{2(n+1)} \left ( {m^2 \over m^2 \, +\  \Box} \right ) ^n \, , 
\label{replace1}
\end{equation}
where only the derivatives contributing to the equation of motion count.    
 We have seen, that when  $m \, \ll \,  L_*^{-1}$, this  de-classicalization  can be understood 
 as a consequence of weakly-coupled UV-completion at the scale $m$.   
  
  $~~~$
  
   What is the meaning of the case  $m \, \gg \, L_*^{-1}$?
  
  $~~~$

    We shall now investigate this limit, and show that in such a case system clasicalizes within a finite energy window, which we shall refer to as the {\it window of classicality}.

   For definiteness,  let us again consider a scattering problem  in a  theory  
   governed by the following equation of motion, 
     \begin{equation}
  \partial^{\mu} \left (\partial_{\mu}\phi \(1\, + \,  2 L_*^4 \, {  m^2 \over  \Box \, + \, m^2} \(\partial_{\mu}\phi\)^2   \) \right ) \,  = \, 0 \, . 
  \label{windeq}
  \end{equation}
 
 We shall again consider an initial state to be a free collapsing spherical wave-packet 
    $\phi_0$, of center of mass energy $\omega$.   From our previous analysis we know, that 
    in $m \, \rightarrow  \, \infty$ limit  scattering classicalizes  for arbitrary $\omega \, \gg \, L_*^{-1}$, with the 
  classicalization radius  $r_*(\omega)$ being given by    (\ref{radstar}). 
  However, for finite (but large)  $m \, \gg \, L_*^{-1}$ the scattering ceases to classicalize above  certain threshold energy $\bar{\omega}$. We shall find this critical energy from the following 
 consideration.   For a free-wave $\phi_0$ the derivative self-interaction in  (\ref{windeq}) 
 gets diminished when the contribution from the $\Box$ operator in the denominator 
 overpowers $m^2$.    For a free collapsing spherical wave we have the following relation $\Box \,\phi_0^2 \, \sim \,  {\omega \over r}\phi_0^2$. 
 Thus, for a given $\omega$  propagator  becomes  dominated by $\Box$ at distances,   
 \begin{equation}
 r \,  \ll  \, \bar{r} (\omega)\, \equiv \,  {\omega \over m^2}  \, . 
 \label{rd}
\end{equation}
We shall refer to $\bar{r} (\omega)$ as the {\it de-classicalization} radius. 
 System classicalizes as long as 
 \begin{equation}
 \bar{r} (\omega) \, \ll \, r_*(\omega) \, ,
 \label{dcond}
 \end{equation}
 and never in the opposite case. 
 Notice, that in contrast to $r_*$, the de-classicalization radius  $\bar{r}$ is a {\it quantum}  length, 
  which diminishes in $\hbar \,  \rightarrow \, 0$ limit.   Indeed, restoring the powers of $\hbar$ in 
  (\ref{rd}), we get,  
 \begin{equation}
  \bar{r} (\omega)\, \equiv \, \hbar^2\,  {\omega  \over m^2}  \,  .
 \label{rdhbar}
\end{equation}

  So any system for which 
  $\bar{r}(\omega)  \,  \gg \, r_*(\omega)$ is quantum, and cannot classicalize. 
  Thus, the equality 
  \begin{equation}
 \bar{r} (\bar{\omega}) \, = \, r_*(\bar{\omega}) \, , 
 \label{obar}
 \end{equation}
defines the upper bound of the classicality window,
 \begin{equation}
     L_*^{-1} \,  < \, \omega  \, < \,  \bar{\omega} \, .
 \label{dcond}
 \end{equation}
  A precise value of $\bar{\omega}$ is model-dependent, but for any system classicality  
  window exist only if (\ref{dcond}) can be satisfied for some $\bar{\omega}  \, > \, L_*^{-1}$. 
    In the example (\ref{windeq}), taking into account (\ref{radstar}),  we have 
 \begin{equation}
 \bar{\omega} \, = \,  L_*^2m^3 \, .
 \label{omegabar}
 \end{equation}
 
 At energies above $\bar{\omega}$ theory stops to classicalize, and becomes  again a  quantum theory.  But, quantum theory of what? 
 Below we shall try to give some insight  for addressing this question. 
 
   \section{Beyond the Classicality window:  Quantum Theory of Big  Objects?} 
  
  At the level of our analysis, the classicalizing theories that in our parametrization correspond to $m \rightarrow \infty$ (equivalently $\bar{\omega} \,   \rightarrow \, \infty$ )  limit,  make a consistent  physical sense. Or, to say it softer, 
  at least at the level of our analysis, we cannot identify anything obviously-inconsistent about such theories.  Their physical meaning can be described as follows. 
  
    At low energies, $\omega \, \ll \, L_*^{-1}$,  
  we deal with a  quantum field theory of a weakly-coupled propagating 
  quantum degree of freedom, $\phi$.  At energies above $L_*$, theory enters a classical
  regime.   Scattering takes place at a macroscopic distance $r_*$ that exceeds all the quantum length-scales in the problem (such as $L_*$).  
  This signals that theory classicalizes.   That is,   high-energy physics is dominated by extended classical field configurations of 
  $\phi$.   As any classical state, these can be viewed as composites of many 
  $\phi$-quanta in a  coherent superpositions.  Because of energy self-sourcing,    
  at higher energies these objects become more and more extended and  thus probe lager and larger  distances.   This tendency continues to arbitrarily high energies.  

 Let us now consider the case of finite $m$ (and thus,  finite $\bar{\omega}$).   
   From the first glance,   a puzzling thing about this  case is,  that theory must become again quantum  above the scale $\bar{\omega}$.   However, quantum theory requires 
existence of corresponding  quantum   degrees of freedom.  By all the accounts the role of such degrees of freedom cannot be played by $\phi$-quanta.  This can be understood by matching some of the properties (such as scattering amplitudes)  of microscopic and macroscopic theories 
at the scale $\bar{\omega}$.  Such matching tells us, that above the scale $\bar{\omega}$, 
the role of propagating quantum degrees of freedom cannot be played by 
$\phi$, since an effective "perturbative" coupling  of such a degree of freedom 
would be strong, enhanced by powers of  $(L_* \bar{\omega})^2$. 
 Thus,  if a sensible theory above $\bar{\omega}$ exists, it must be a quantum theory of something else.   
 
  We would like to suggest one such possibility. 
 The hint comes from thinking about the properties of classicalon  configurations  as of  functions 
 of a continuous parameter $\omega$.  For  values of $\omega$ within the classicality window, 
 classicalon configurations (by default) are characterized by the size $r_*(\omega)$ and energy $\omega$. 
 Within this window their dynamics is governed by classical physics, and quantum influence is
 negligible.  What happens with classicalons  above the classicality window? 
 The fact that theory de-classicalizes means that classicalons  of mass above $\bar{\omega}$ 
 can no longer exist.   In other words, even if we try to prepare such configurations, they should become so quantum-mechanically-unstable, that it should not make sense to talk about them as 
 of well-defined states. In other words,  life-time of (would-be) classicalons  must become shorter than their size.  In order to understand better what picture we have in mind, we shall develop  the following analogy. 
  
    We shall speculate, that perhaps a QCD-type theory can be considered  as a limiting case 
   of such a situation.  For this,  let us first consider a confining QCD-type theory  in which quark masses, $m_q$,  are higher  than the QCD scale, $\Lambda_{QCD}$.  As we know, because of confinement such a theory contains extended objects, the QCD electric  flux tubes.  The flux-tubes can be either closed or end on heavy quarks, but otherwise they cannot terminate.    A long 
   QCD-tube, can break quantum-mechanically by nucleating 
   quark-anti-quark pairs.  But, since quarks are heavy, the process is exponentially-suppressed. 
 Nevertheless,  the non-vanishing  probability of breaking limits the possible size of classical strings 
 and creates a scale somewhat analogous to the  upper bound of classicality window, $\bar{\omega}$.  
    
 The rate of string-breaking can be estimated, e.g., by treating 
 the quark-anti-quark production as  a Schwinger pair-creation process \cite{pair1} in one-dimensional electric field  of the string, or by dualizing decay of unstable cosmic strings (magnetic flux-tubes)  via creating monopole-anti-monopole pairs \cite{pair2}.  The resulting 
 probability of breaking per unit length per unit time is given by, 
 \begin{equation}
 \Gamma \, \sim  \,  \Lambda_{QCD}^2 \, {\rm e }^{-c (m_q/\Lambda_{QCD})^2} \, ,  
  \label{ratebreak}
  \end{equation} 
 where $c$ is a model-dependent  numerical factor of order one. 
  The above rate can be estimated as follows.  By removing a 
  portion of string of length $L$ the energy increment is 
  \begin{equation}
  \Delta E \, \sim \,  2 m_q \, - \,  \Lambda_{QCd}^2 \, L \,, 
  \label{deltaE}
  \end{equation}
 where the first term is the positive price that comes from nucleating quark-anti-quark pair, 
 whereas the second term is the gain in energy due to removing the  portion of a  string 
 of length  $L$ and tension $\Lambda_{QCD}^2$.   Thus, the size of a critical "bubble" that one has to nucleate  is given by 
 $L_c \, \sim \,  m_q/\Lambda_{QCD}^2$ and the Euclidean action on a bounce 
 is $S_E \, \sim \, (m_q/\Lambda_{QCD})^2$.   Thus, the lifetime of a flux tube of the 
 length $L\, \gg \, \Lambda_{QCD}^{-1}$ is
 \begin{equation}
 \tau_L \, \sim \,  {1 \over \Lambda_{QCD}^2 L} \,  {\rm e }^{c (m_q/\Lambda_{QCD})^2}  \, .
 \label{lifetime}
 \end{equation}
  As long as this quantum lifetime is much longer than the string length, $L$, string can be considered to be  a well-defined classical object.  Once the lifetime becomes shorter than $L$, the QCD string  ceases  to be a well-defined classical state and its dynamics is governed  by quantum mechanics.  Thus, the classicality condition is $\tau_L \, \gg \, L$. This condition defines a critical 
  length scale, 
  \begin{equation}
   \bar{L} \, \equiv\,  \Lambda_{QCD}^{-1} \, {\rm e }^{{c\over 2} (m_q/\Lambda_{QCD})^2} \, ,
   \label{lcrit}
   \end{equation}
   beyond which the classical strings cease to exist.   The corresponding string 
   mass,  
 \begin{equation}
 \bar{\omega} \, = \,   \bar{L} \Lambda_{QCD}^2 \, , 
 \label{massupper}
 \end{equation}
 sets an  upper bound on classicality window (\ref{dcond}), which in this case becomes,
  \begin{equation}
     \Lambda_{QCD} \,  < \, \omega  \, < \, \bar{\omega} \, = \,   \bar{L} \Lambda_{QCD}^2 \, .
 \label{windowQCD}
 \end{equation}
 
  In order to understand how far the analogy goes, let us consider  a high-energy scattering process
  in such a theory.  Because of confinement,  at high energies  we can distinguish 
  two different types of observers.   The first is a short-distance observer that resolves physics 
  at distances shorter than the  QCD length,  $L_{QCD} \, \equiv \, \Lambda_{QCD}^{-1}$. 
  This observer describes the collision process in terms of quarks and gluons and can probe 
  arbitrarily short distances.   The second one is a long-distance observer that can probe distances 
  $\gg \, L_{QCD}$. The latter observer describes scattering process in terms of 
  QCD-flux tubes  (glueballs and hadrons).  In order to connect the two observations, one has to go 
  through the process of "hardonization" that takes place at distance $L_{QCD}$. However, 
  the long-distance observer can in principle integrate out this complicated  dynamics and describe 
  the scattering process in terms of an effective theory of interacting QCD flux-tubes.  This description then  should  be almost classical within the energy window  given by  (\ref{windowQCD}) and the corresponding distances  $L_{QCD} \,  < \, L\,  <  \bar{L}$. 
 
  Of course, as we noted, the strings can break by quark-anti-quark production, but for the 
  strings within  the above window this process is negligible. 
  Also, even without quark production, vibrating long strings can decay into smaller loops. 
  First, an oscillating string loop can shorten and quantum-mechanically radiate smaller loops
  (glueballs). Secondly, a vibrating long string   
  can  chop off  smaller loops
  every time the flux-tube intersects with itself.   The latter process is very similar to inter-commutation of oscillating cosmic string loops \cite{strings} and is described by similar dynamics. 
  
  In the language of  QCD resonances  both processes are seen  
  as decay into lighter glueballs. However,  for  a flux tube of length  $L \, \gg L_{QCD}$  
    and curvature radius $\sim L$, the decay into very small loops,   of size $\ll L$, is suppressed. 
    Moreover, a  characteristic  vibration period  of such a string is $\sim L$, and therefore the lifetime  is at least as long as  $L$.    So dynamics of such a  macroscopic string-loop is well-approximated by classical physics at least on the time-scale $\sim L$.  
    
    The most important fact about  the above  analogy is the finiteness of classicality window. 
 Since no well-defined classical strings exist with mass above $\bar{\omega}$ and length above 
 $\bar{L}$, 
 the long distance observer at energies  $\omega \, \gg \, \bar{\omega}$ is  forced back to quantum description 
 and the effective theory at those distances is a theory of breakable  flux-tubes. 
   
  Let us now carefully go through the valid analogies  as well as  their limitations. 
    
  \subsection{Analogy}  
    
   We tried to establish a certain analogy between the theories with finite classicality
   window and QCD-type theory with breakable flux tubes. 
  The essential property of any former theory is the existence of at least three length-scales. 
  Two of them, $L_*$ and $\bar{r}$  are quantum in nature and vanish in 
 the limit  $\hbar \rightarrow 0$. Whereas the third one $r_*$ is classical and independent 
 of $\hbar$.   Another essential 
 property is that two of the scales, $\bar{r}(\omega)$ and $r_*(\omega)$ are energy-dependent. 
 The classicality window exist if for some interval of energies  (\ref{dcond}), the scale   
  $r_*(\omega)$ can be a dominant length scale.  For  such energy  the classicalon configuration 
  is formed before the system has any chance to probe the quantum length-scales $\bar{r}$ or 
  $L_*$.  
  
  In case of QCD flux tubes,  the roles of the scales  $L_*$, $r_*$ and $\bar{r}$ are  played 
  by the  QCD length $L_{QCD}$,  the string length 
  $L$, and the ratio of the latter length-squared to the critical length ($L^2/\bar{L}$), 
     respectively.   
  In  other words, the dictionary is: 
  \begin{equation}
   L_* \, \rightarrow  \,  L_{QCD} \, 
  \label{starqcd}
   \end{equation}
  \begin{equation}
   r_* \, \rightarrow  \,  L \, 
 \label{stringlength}
   \end{equation}
  \begin{equation}
  \bar{r} \, \rightarrow  \,  {L^2 \over \bar{L} }\, 
  \label{rdlife}
   \end{equation}
   The physical meaning of  the above correspondence is clear.  Obviously the QCD-length, 
   $L_{QCD}$,  plays the same role as $L_*$, since it marks the length scale beyond which classical 
   flux tubes exist.  Both of the scales are quantum in nature.   The string length, $L$, is an obvious classical scale in the problem.  We can talk about classical strings only as long as $L > L_{QCD}$. 
   Finally, it is obvious that the ratio $L^2/\bar{L}$ controls  classicality of 
   strings,  since strings cease to exist when  $\bar{L} \, < \, L$. 
   
   \subsection{Limitations:  Light-quark  QCD as a Collapsed Classicality Window?}
   
      The above analogy has to be taken with an extreme care, as it does have obvious limitations. 
   At the current level of the analysis it only serves as  an existence proof of a situation when  a classical behavior  within a limited interval of energies is possible.  It shows, that  we can have a well-defined microscopic theory for which the effective high-energy/long-distance behavior 
   can be understood as interpolation between quantum to classical and back to quantum regimes. 
   The very last  transition being triggered by quantum instability of the extended (would-be) classical objects that sets-in above their  critical size and energy.  
   
    On the other hand there are clear differences.   For example, in high-energy  QCD 
    we can scatter heavy quarks   at arbitrarily short distances.  In this respect, the
     roles of $r_*$  and $L$ are very different.   In the example  (\ref{windeq}) the scattering does 
     happen at distance $r_*$,  whereas in QCD the long tubes are an outcome of  a 
     short-distance scattering process through hadronization. 
     
      This difference has to do with the  qualitative differences between the low energy spectra of the two theories.     The model  (\ref{windeq}) at low energies,  $\omega \ll L_*^{-1}$,  includes  
a light propagating degree of freedom, a Goldstone boson $\phi$.   Whereas, QCD 
without light quarks is a theory with the mass gap, with no propagating degrees of freedom 
in deep infrared  $\omega \ll \Lambda_{QCD}$. 
 Because of this, classicalization  of QCD flux tubes  does not represent an UV-completion 
 of any low-energy Goldtone-type  theory.   
 
 We could have tried to make the analogy closer by 
 taking the limit of  light-quarks, $m_q \, \ll \, \Lambda_{QCD}$. Of  course, in such a case
 there is a well-defined low-energy description in terms of a pion chiral Lagrangian, but now 
 classicalization  is spoiled from the other end, since QCD flux tubes simply become unstable. 
  All the three length-scales, classical and quantum, become of the same order,
 \begin{equation}
 L_{QCD} \, \sim L\,  \sim \, \tau_L \, .
  \label{colapse!}
   \end{equation}
  The classicality window  collapses to a point!
  
  We thus see,  that at least in ordinary QCD,  there is no parameter choice 
 for which  one could have a low energy  pion theory and simultaneously maintain a 
 significant classicality window.

 This consideration indicates that in certain crude (but well-defined) sense, ordinary 
 QCD can be thought of as a limit of would-be classicalizing theory  in which the classicality  
 window collapsed to a single scale, $\Lambda_{QCD}$.  Whether the above-presented  view sheds any useful light  at the QCD-confinement  dynamics, is unclear, but it  is certainly  illustrative  for understanding regimes of  classicalization phenomenon.

   \section{conclusions}
   
     Many weakly coupled microscopic theories, such as  models with spontaneously broken 
     global symmetries,  Higgsed gauge theories  and  high-energy QCD, are known to result at low energies in derivatively interacting massless  or light  Nambu-Goldstone type degrees of freedom. 
  On the other hand, it was argued recently \cite{class1, class2}, that  such theories have a tendency to classicalize  in very high energy scatterings.  Namely,  the theories in question contain a classical  length-scale $r_*$ that governs the range of  interaction and grows with increasing center of mass energy in such a way  that  exceeds all the quantum length-scales in the problem.   Classicalization is therefore  taking place  at  distances  much larger than the Compton wave-length  of any weakly coupled physics that would-be responsible  for maintaining the perturbative unitarity at short distances.  
 
  This behavior  creates a seeming puzzle, since a long-distance theory should not know 
  about the short-distance UV-completion and thus,  should  continue to classicalize even when 
  the short distance interactions are UV-completed by weakly-coupled physics that maintains perturbative unitarity.  But,  how can these two dramatically 
  different behaviors of the scattering amplitudes be reconciled? 
  
    In the present note we have resolved this puzzle, and have shown that the two behaviors 
    are not reconciled,  and the theory can chose only one way:   Classicalize or not  classicalize. 
    Integration-in of any weakly-coupled short distance physics that maintains perturbative unitarity automatically kills clasicalization:  $r_*$-radius shrinks below the quantum-mechanical length-scales and becomes irrelevant; system de-classicalizes. 
    
       The reason why  we arrived to a seeming puzzle to start with was the wrong intuition, that 
 the high-energy structure of the theory should not affect its  long-distance behavior. 
 In classicalizing theories this intuition breaks down, since  high-energy physics probes long distances, due to existence of the extended classical objects. 
  In other words, in classicalizing theories very high energies are not equivalent to short   
 distances but rather to very large distances.  As a result, 
 the  energy densities  corresponding to  $r_*$ are absolutely enough for theory to ``know" 
 whether it classicalizes or not.  This is why a  perturbative-unitarity-restoring  weakly-coupled physics with mass $m$ below 
 $L_*^{-1}$   ruins classicalization, which, by the way, in this case is no longer needed. 
 On the other hand, if we deprive  a  theory of a weakly-coupled UV-completion 
 by pushing $m \rightarrow  \infty$,
 theory puts up a 
 "self-defense" and unitarizes by classicalization.    
 
  We have shown, that  interpolation between the two regimes can be parameterized by defining a 
  new quantum  radius $\bar{r}$,  which we called the radius of  de-classicalization. 
  System de-classicalizes when there is no choice of center of mass energy for which 
  the classical radius $r_*$ could exceed the quantum scale $\bar{r}$, and classicalizes in the opposite  case.    The  former  situation takes place when  $m \, < \, L_*^{-1}$, whereas the  
  latter, when $m \, = \, \infty$.    

     This consideration lead us to a logical possibility for the existence of an intermediate regime, in which   $r_*$ dominates over $\bar{r}$ only within a finite energy window, called window of classicality. 
 This phenomenon takes place when $m$  is  large but finite,
 $m \, \gg \, L_*$.    However,  in this case $m$  can no longer
 be identified with a mass scale  of new  weakly-coupled quanta and question arises about its physical meaning.   In other words,  finiteness of the classicality 
 window rises the question about the  physical meaning of theory beyond this window. 
 We have given an evidence that above the classicality window we deal with a theory of unstable 
 extended objects.
  
  In order to support this intuition, we  have established an analogy with QCD-type theory that 
  contains such objects, the unstable  QCD flux tubes.   However,  for flux  tubes of size 
  $\gg \, L_{QCD}$ to exist,  quarks have to be heavier than the QCD scale.  But,  in this case no low energy  analog of Goldstone-type classicalizer field  $\phi$ exists.   Creating such an analog, by taking the light quark limit, automatically makes flux-tubes unstable and 
  collapses the classicality window to  a point. 
 
  This fact sheds for us two different lights.  First, it provides  for us an alternative language, 
 in terms of QCD flux tubes,  for  explaining  why 
  theory of real QCD pions  does not classicalize, despite the presence of derivative self-interactions.  
   Secondly,  it  tells us that  real QCD can be viewed as a theory with a collapsed classicality window. 
   
    Finally it wold be interesting to explore a possible connection (if any) between some of the presented ideas about classicalization and recently suggested possible UV-finiteness  of 5D super-Young-Mills \cite{5d}.

  \vspace{5mm}
\centerline{\bf Acknowledgments}
We thank  Cesar Gomez and Akaki Rusetsky for many useful discussions. It is  a pleasure to  acknowledge valuable discussions on classicalization  with 
Gian Giudice, Alex Kehagias and David Pirtskhalava.  
We thank Luis Alvarez-Gaume,  Gerhard Buchalla  and Martin L\"uscher for asking questions 
(hopefully) addressed   in this work. 
 This work was supported in part by Humboldt Foundation under Alexander von Humboldt Professorship,  by European Commission  under 
the ERC advanced grant 226371,  by  David and Lucile  Packard Foundation Fellowship for  Science and Engineering, and  by the NSF grant PHY-0758032. 

 \vspace{5mm}
\centerline{\bf Note Added}

  Before submitting this paper,  a paper by Heisenberg \cite{Heisenberg} from 1952 was brought 
  to our attention 
  \footnote{We thank Cesar Gomez for  bringing \cite{Heisenberg}  to our attention.}. In this paper Heisenberg tries to  account for high-multiplicity in strong interaction scattering 
  by studying propagation of  shock-waves in pion field, and argues that derivative self-interactions 
  are important for the correct description of the scattering.  Since QCD-type theories only serve as an useful rough analogy, rather than the  main focus of our work,  the phenomenological validity  of Heisenberg's result  is  secondary for this comment. What  is remarkable for us is his appreciation  of importance of the derivative self-couplings for creating multiplicity of states in high-energy scattering.  This emphases resonates with our  ideas on  classicalization.

\end{document}